# Trapped Water on Silicates
# in the Laboratory and in Astrophysical Environments

Alexey Potapov[1,2], Cornelia Jäger[1], Harald Mutschke[3], Thomas Henning[4]

[1]*Laboratory Astrophysics Group of the Max Planck Institute for Astronomy at the Friedrich Schiller University Jena,*
*Institute of Solid State Physics, Helmholtzweg 3, 07743 Jena, Germany*
[2]*Institute of Geosciences, Friedrich Schiller University Jena,*
*CEEC II, Lessingstr. 14, 07743 Jena, Germany, e-mail:* alexey.potapov@uni-jena.de
[3]*Astrophysical Institute and University Observatory, Friedrich Schiller University Jena,*
*Schillergässchen 2-3, D-07745 Jena, Germany*
[4]*Max Planck Institute for Astronomy, Königstuhl 17, D-69117 Heidelberg, Germany*

**Abstract**

Existence of strongly bound water molecules on silicate surfaces, above the desorption temperature of water ice, has been first predicted by computational studies and recently demonstrated by laboratory experiments. Such trapped water may be present in various astrophysical environments and there is now evidence for its presence in the diffuse interstellar medium and in extraterrestrial particles. We present here new results of a laboratory study of the phenomenon of trapping (strong bonding) of water molecules by silicates. We show that the efficiency of trapping is strongly dependent on the properties and composition of the surface. Our results point out that the presence of trapped water should be due to the hydrophilic properties of the silicate surface and that the nature of trapping is physical (physisorption rather than chemisorption). We demonstrate that water can be trapped on silicates up to the temperatures of about 470 K, which speaks for the presence of wet silicate grains in the terrestrial planet formation zone in planet-forming disks. Studying the thermal and UV stability of trapped water, we conclude that the detection of trapped water in the diffuse ISM speaks for its efficient continuous formation. We discuss our results as relevant to fundamental scientific questions, such as the oxygen depletion problem, the origin of water on Earth, and the formation of rocky planets.

1. **Introduction**

Water is one of the main molecular species in different astrophysical environments, such as prestellar cores, protostellar envelopes, planet-forming disks, and planetary atmospheres, and it is evidently required for the formation of life as we know it. Understanding the role of water in planet formation and chemistry of protoplanetary disks and its delivery to rocky planets is a key requirement for understanding the atmospheric composition of planets and the origins of life on Earth and, potentially, on extrasolar planets.

It is typically assumed that solid-state water in astrophysical environments exists only in cold dense regions, such as molecular clouds, protostellar envelopes and planet-forming disks beyond the water snowline, in the form of molecular ice. Solid-state water is a part of dust grains, which are typically considered to consist of mainly carbon- and silicate-based particles and molecular ices. Water is the main constituent of these ices accounting for more than 60% of the ices detected in most lines of sight (Whittet 2003; van Dishoeck et al. 2021). Evidence to date shows that ice is physically mixed with dust in space. The examination of the samples returned by the Stardust mission showed that the dust and water-ice agglomerates were mixed before comets formed in the outer solar system (Brownlee et al. 2006). Results of the Rosetta mission to the comet 67P/Churyumov–Gerasimenko demonstrated that the core consists of a mixture of ices, Fe-sulfides, silicates, and hydrocarbons (Fulle et al. 2017). Different models of comets, such as dirty snowball, icy glue, and fractal aggregates (A'Hearn 2011) mean that ice is mixed with dust in cometary nuclei. A recent review paper on the formation of comets (Blum et al. 2022) considers dust/ice mixtures as a reliable model for building blocks of pristine comets. Moreover, dust grains in prestellar cores and planet-forming disks are believed to form fractal aggregates characterized by a very high porosity. This is suggested by the analysis of cometary dust particles (Harmon et al. 1997; Fulle et al. 2000; Hörz et al. 2006; Fulle & Blum 2017), dust evolution models (Ossenkopf 1993; Kataoka et al. 2013; Tazaki et al. 2016), laboratory experiments on the dust particle aggregation by collisions (Wurm & Blum 1998; Krause & Blum 2004), and on the gas-phase condensation of grains with their subsequent aggregation on a substrate (Jäger et al. 2008; Sabri et al. 2014). It is reasonable to assume that ice fills the pores of grain aggregates. Moreover, the dust/ice mixing should occur when nm-sized/submicron grains containing both ice and dust grow into larger aggregates in the ISM and in planet-forming disks. Very recently, by comparing laboratory IR data on dust/ice mixtures and observational data, it has been demonstrated that a substantial fraction of water ice may be mixed with silicate grains in protostellar envelopes and protoplanetary disks (Potapov et al. 2021).

In the laboratory experiments mimicking physico-chemical processes in cold astrophysical environments, pure water ice thermally desorbs completely at 160 – 180 K (Collings et al. 2004; Bolina et al. 2005; Smith et al. 2011). However, in the majority of laboratory experiments ice is deposited onto standard laboratory substrates, which are not characteristic of cosmic dust grains. More realistic laboratory experiments on silicate-grains/water-ice mixtures have demonstrated that a considerable part of water molecules mixed with silicate grains at low temperatures is trapped (strongly bound) on the grains at temperatures exceeding the desorption temperature of water ice, up to at least 200 K (Potapov et al. 2018a; Potapov et al. 2018b; Potapov et al. 2021). This finding reinforces the results of the calculations on the interaction of water with silicates showing the presence of stronger adsorption sites on the surface, where water molecules may be retained at high temperatures (Stimpfl et al. 2006; Muralidharan et al. 2008; King et al. 2010).

These experimental and modelling results suggest that water molecules can be trapped in refractory dust grains and may be present in the solid state in astrophysical environments, where we do not expect it, e.g., in the diffuse interstellar medium (ISM), in planet-forming disks inside the snowline, and in the atmospheres of exoplanets. Based on the combination of laboratory data and infrared observations, evidence of the presence of solid-state water in the diffuse ISM has been provided (Potapov et al. 2021). Furthermore, a recent comparison of infrared spectra of laboratory-made refractory organic residues on silicates to the spectra of cometary particles returned by the Stardust mission, interplanetary dust particles, and meteorites speaks for the presence of trapped water in these extraterrestrial particles (Potapov et al. 2022). We note that trapped water originates from silicate/ice mixing at low temperatures and has no relation (maybe only as a possible precursor) to phyllosilicates that also present a great interest as possible solid-state water reservoirs at high temperatures as they can retain water when heated up to a few hundreds of degrees centigrade, e.g., (Davies 1996; Beck et al. 2014; D'Angelo et al. 2019; Thi et al. 2020) and a review (Elmi et al. 2016).

The presence of trapped water on silicates in various astrophysical environments may have a number of important consequences. Detection of trapped water in the diffuse ISM contributes to solving the oxygen depletion problem providing an earlier unaccounted location of oxygen. Detection of trapped water in planet-forming disks (particularly, inside the water snowline) would lead to a better understanding of the origin of water on Earth and terrestrial planets and of the formation process of rocky planets. Detection of trapped water in exoplanet atmospheres would give new information regarding the formation and chemistry of atmospheric clouds.

Appropriate observations are on the horizon with the launch of the James Webb Space Telescope (JWST). For more reliable modelling of astrophysical processes, prediction of and comparison to upcoming observations, it is important to understand better the properties of trapped water including the nature of trapping, physisorption (meaning Van der Waals or hydrogen bonding) or chemisorption (meaning formation of covalent or ionic bonds) on silicates, the first with the adsorption energy of about 600 K (Stimpfl et al. 2004).

The goals of the present study were to investigate the amount and stability of trapped water in an extended temperature range (as compared to previous studies) and on various surfaces, and to provide motivation for a further search for trapped water in various astrophysical environments.

## 2. Methods

**Preparation of silicate and silicate/kerogen samples**

Amorphous $MgSiO_3$ silicates were produced in a setup described in detail elsewhere (Sabri et al. 2014). In brief, the production of highly-porous deposits of nanometre-sized amorphous $MgSiO_3$ grains was achieved by pulsed laser ablation of a MgSi 1:1 target, subsequent condensation of the evaporated species in a quenching atmosphere of 4 mbar $O_2$, adiabatic extraction of the condensed grains from the ablation chamber through a nozzle and a skimmer, and their deposition onto a KBr substrate fixed in a separate (deposition) chamber. The thickness of the silicate grain deposits was 100 nm and was measured by a quartz crystal resonator microbalance using known values for the KBr deposit area of 1 $cm^2$ and silicate density of 2.5 g $cm^{-3}$. An important parameter of the structure of the amorphous silicate deposits is their very high porosity and a corresponding large surface area, which is a few hundred times larger than the surface of a KBr substrate (Potapov et al. 2020). The composition, structure and morphology of silicate grains produced in our experiments were characterized in a plenty of studies. The interested reader can find details in, e.g., (Jäger et al. 2008; Sabri et al. 2014; Jäger et al. 2016). Energy-dispersive X-ray (EDX) spectroscopy analysis of the grain ensembles revealed Mg to Si ratios between 0.95 to 1.1. The measurement of a large sample area is $Mg_{0.98}SiO_3$. EDX of amorphous silicate grains is not trivial, because small silicate grains can be easily damaged by the electron beam during the measurement. We choose the EDX parameters to minimize this damage, but not to achieve the highest accuracy. In principle, the final composition of the siliceous condensate is determined by the laser target composition. Very small deviations from the $MgSiO_3$ composition are absolutely not relevant for the structural and spectral properties, as well as other characteristics of this material. The characteristics of amorphous silicates is that within a small sub-nm or nm scale derivations of the amorphous structure may happen.

In order to modify the morphology and structure of the condensed silicates (in particular the grain sizes and porosity), two samples were annealed in a two mbar argon atmosphere for three hours at 500 and 900°C, respectively. It is now well accepted that the refractory organic materials in cometary dust as well as in meteorites are dominated by high molecular weight organic components very similar to the kerogen-like material (Osawa et al. 2009; Matthewman et al. 2013; Wooden et al. 2017). The disordered structure of kerogen is not well-defined and can be described as a mixture of aromatic and aliphatic sub-units containing a large number of functional groups. To cover amorphous silicate grains by a kerogen-like material, two silicate samples were cooled down to 10 K and $CH_3OH$ ice was deposited onto the silicates. The $CH_3OH$ ice thickness was calculated from its vibrational band at 1026 cm$^{-1}$ using the band strengths of $1.8 \times 10^{-17}$ cm molecule$^{-1}$ (Hudgins et al. 1993) and was about 200 nm considering 1 cm$^2$ of the KBr deposition area. After deposition, the silicate/ice mixtures were irradiated for 1 or 5 hours by a broadband deuterium lamp (L11798, Hamamatsu) with a flux of $10^{15}$ photons cm$^{-2}$ s$^{-1}$. The lamp has a spectrum from 400 to 118 nm with the main peaks at 160 nm (7.7 eV) and at 122 nm (10.2 eV). After irradiation, the cooling was stopped and the samples were warmed up to room temperature. UV photons cause carbonization processes of the ices that lead to the formation of a refractory material, the best analog for which is kerogen. As obtained in our experiments from measurements of IR spectra of kerogen-like materials comparing the intensities of CH bands, the amount of the material increases with the UV irradiation time. The purpose of irradiating the silicate/ice mixtures for 1 or 5 hours was to produce two samples with different amounts of the kerogen-like material.

**Deposition of water and low-temperature (200 – 300 K) measurements**

Amorphous and crystalline silicate and silicate/kerogen samples were cooled down to 10 K and water was deposited onto them from a water reservoir through a leakage valve and a capillary tube. The depositions were performed in a high vacuum chamber with a base pressure of $5 \times 10^{-8}$ mbar at 10 K. The thickness of the water ice was calculated from the 3.1 μm band area using the band strength of $1.9 \times 10^{-16}$ cm molecule$^{-1}$ (Mastrapa et al. 2009) and was about 200 nm calculated for the 1 cm$^2$ KBr surface using the density of 1.1 g cm$^{-3}$ for high-density amorphous water ice (Narten et al. 1976). Thus, the $MgSiO_3/H_2O$ mass ratio in the samples after $H_2O$ ice depositions was about 1.25.

After the $H_2O$ deposition, the samples were heated up consequently to 200, 250 and 300 K and their IR spectra were measured in the transmission mode using a Fourier transform infrared (FTIR) spectrometer in the spectral range of 6000 – 600 cm$^{-1}$ at a resolution of 1 cm$^{-1}$. The spectra of pure KBr substrates recorded at 200, 250 and 300 K were used as reference spectra. To check the stability of trapped water with time, IR spectra of a MgSiO$_3$/H$_2$O sample were measured at 200 K during 8 hours. To check the stability of trapped water under UV irradiation, IR spectra of a MgSiO$_3$/H$_2$O sample were measured at 200 K during 3 hours of UV irradiation using the deuterium lamp described above.

**High-temperature (300 – 700 K) measurements**

After the low-temperature measurements the samples were extracted from the deposition chamber and placed into an external high-temperature high-pressure (HTHP) cell described in detail elsewhere (Zeidler et al. 2013). With this cell, it is possible to heat samples to temperatures up to 1073 K and to take IR spectra at the same time. The HTHP cell is supplied by an external temperature controller by which the heating temperature and the temperature gradient can be set. The controller measures the temperature of the sample holder and the body of the cell. The spectra were measured in vacuum (< 10$^{-3}$ mbar). The samples were heated up consequently to 323, 373, 423, 473, 523, 573, 623 and 673 K (50, 100, 150, 200, 250, 300, 350 and 400°C) with the heating rate of about 2 degrees per minute and their IR spectra were measured in the transmission mode using an FTIR spectrometer in the spectral range of 6000 – 400 cm$^{-1}$. The spectra of KBr substrates recorded at the same temperatures were used as reference spectra. Two samples have been measured showing reproducible results.

3. **Results**

Figure 1 presents IR spectra of two samples: pure H$_2$O ice and H$_2$O ice mixed with MgSiO$_3$ grains at 10 K. The spectra have been taken at 200 K, above the desorption temperature of H$_2$O ice. In the case of pure H$_2$O ice, there are no spectral features observable that indicates that the ice is completely desorbed. In the case of H$_2$O ice mixed with MgSiO$_3$ grains, O-H stretching and O-H-O bending features of H$_2$O molecules are clearly visible. The stretching band of trapped water is observed in the range of 3750 – 3000 cm$^{-1}$ (2.67 – 3.33 μm). The bending band of trapped water is observed in the range of 1750 – 1530 cm$^{-1}$ (5.71 – 6.54 μm). The band around 1000 cm$^{-1}$ (10 μm) corresponds to Si-O stretching vibrations in the silicates.

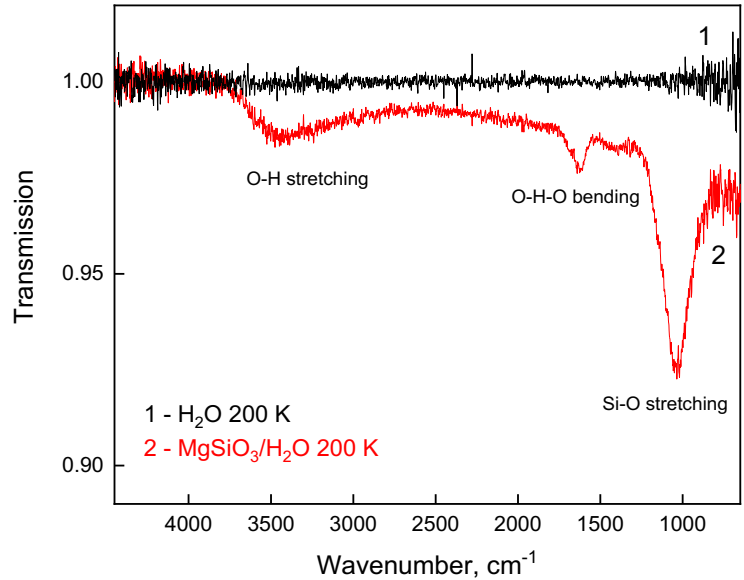

Figure 1. Transmission spectra of pure $H_2O$ ice (1) and $H_2O$ ice mixed with silicate grains at 10 K (2) at 200 K, above the desorption temperature of $H_2O$ ice.

A temperature dependence between 200 and 300 K of the amount of remaining water (in percent of the deposited amount) in a silicate/$H_2O$ sample is presented in Figure 2. The amount of trapped water decreases with the temperature by a factor of about 2 for 250 K and of about 10 for 300 K, as compared to 200 K. Such an efficient temperature desorption tells us that at least a part of trapped molecules, which desorb, are physisorbed and not chemisorbed on the surface of silicates. We note that the "thickness" of the trapped water was calculated in the same way as for the amorphous water ice, from the 3.1 µm band area using a band strength of $1.9 \times 10^{-16}$ cm molecule$^{-1}$. As the band strength of trapped water is not known, the presented values should not be taken at face values.

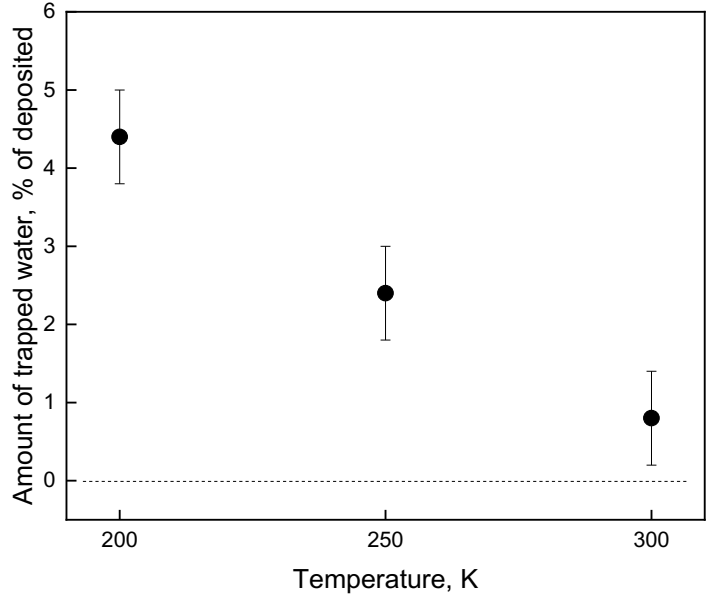

Figure 2. Temperature dependence between 200 and 300 K of the amount of remaining water in a silicate/$H_2O$ sample as a percentage to the deposited amount. The errors reflect statistical variations.

The next question is the stability of trapped water. As it was recently demonstrated (Potapov et al. 2021), solid-state (trapped) water is present in the diffuse ISM. The temperature in diffuse clouds is about 80 K (Tielens 2013). In this case, water molecules should survive (i) lifetime of a cloud and (ii) irradiation by UV fields that might lead to efficient water photodesorption. In the next figure we present dependences of the amount of remaining water in the silicate/$H_2O$ samples at 200 K on time (the sample was kept at 200 K for 8 hours) and on a UV fluence.

A slight decrease of the amount of trapped water can be observed in Figure 3 (left) after keeping the sample at 200 K for 8 hours. The slope of the linear fit is $0.034 \pm 0.033$ hour$^{-1}$. Even considering the lowest slope value of $0.001$ hour$^{-1}$, on the timescale of millions of years (the lifetime of a cloud) trapped water should completely desorb. A greater decrease of the amount of trapped water can be observed in Figure 3 (right) after 3 hours of irradiation providing a final UV fluence of $9\times10^{18}$ photons cm$^{-2}$. The slope of the linear fit is $0.1 \pm 0.05$ $\times10^{-18}$ cm$^2$ photon$^{-1}$. It corresponds to a photodesorption yield of about $0.2\times10^{-3}$ molecules photon$^{-1}$, which is one order of magnitude lower than the values obtained for the photodesorption yield of water ice at 10 K (Westley et al. 1995; Öberg et al. 2009). However, these values show that after the fluence of about $10^{20}$ or even earlier, all trapped $H_2O$ molecules should be UV-sputtered from the dust surface. In diffuse regions of the ISM, the UV fluence is about $10^{22}$ photons cm$^{-2}$ (Moore 1999). Thus, no trapped water should be present in diffuse clouds if we consider only destruction processes. Obviously, the detection of solid-state water in the diffuse ISM speaks for its efficient continuous formation. This experimental result reinforces the conclusion made by Cuppen and Herbst (2007) on the basis of their model that in diffuse – translucent clouds water desorption and destruction by photons is balanced by its formation. The formation goes most probably through reactions between hydrogen and oxygen atoms/molecules on the dust surface as it has been shown in a number of experiments, e.g., (Ioppolo et al. 2008; Oba et al. 2009; Jing et al. 2011; Accolla et al. 2013).

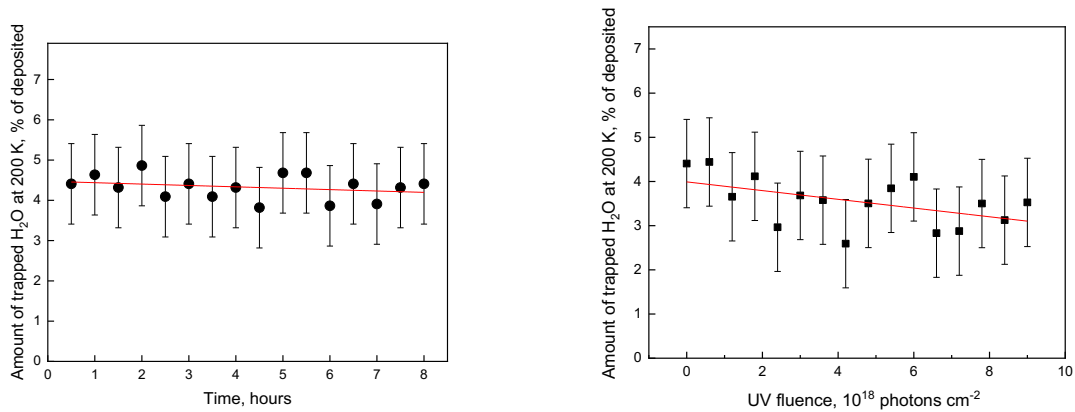

Figure 3. Left: Time dependence of the amount of remaining water in a silicate/H$_2$O sample at 200 K as a percentage to the deposited amount and its linear fit. Right: Dependence of the amount of remaining water in a silicate/H$_2$O sample at 200 K on the UV fluence as a percentage to the deposited amount and its linear fit.

We have also studied the stability of trapped water at temperatures higher than 300 K. In Figure 4, we present a temperature dependence of the amount of remaining water in a silicate/H$_2$O sample.

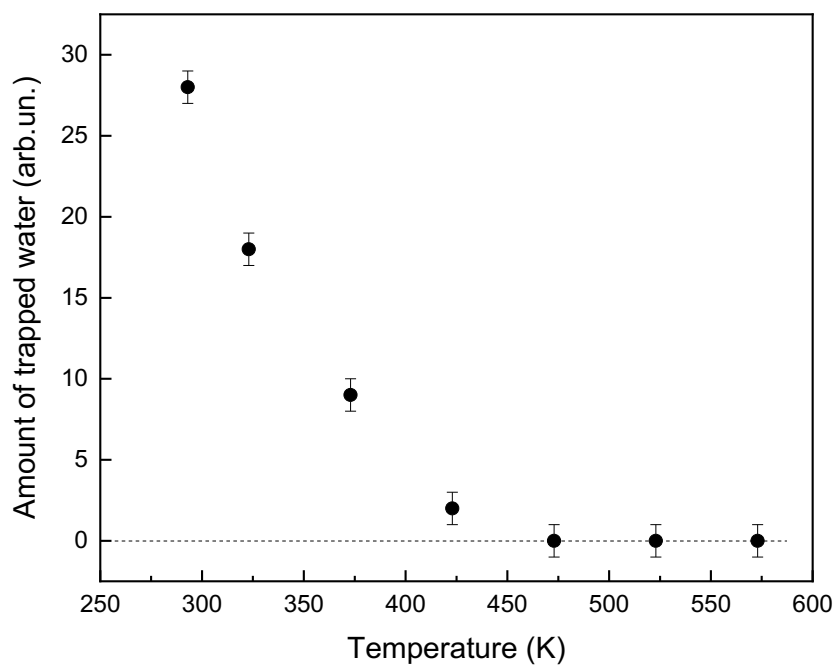

Figure 4. Temperature dependence between 293 and 573 K of the amount of remaining water in a silicate/$H_2O$ sample.

As can be seen from Figure 4, trapped water desorbs completely at about 470 K. Note that trapped water in these samples is a mixture of $H_2O$ molecules originated from the ice and molecules adsorbed on silicate grains from the atmosphere due to exposure of the samples to air. That is why arbitrary units are provided on the y-scale of Figure 4 as they cannot be directly linked to the units in Figures 2 and 3. We have to admit that for now we do not see a possibility to distinguish between these two types of trapped water molecules in our silicate samples.

The next important question is the efficiency of trapping of water on silicates having different structures (amorphous and crystalline), as both have been observed in the ISM, and on silicates mixed with refractory hydrocarbons, a product of energetic processing of C- and H-bearing ices in interstellar and circumstellar environments. Figure 5 shows the typical structure of the silicate grains after gas-phase condensation, deposition, and subsequent thermal annealing.

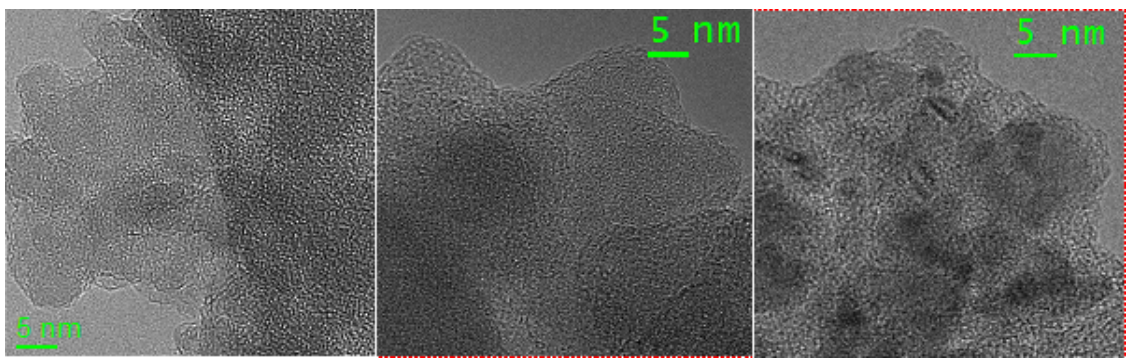

Figure 5. High-resolution transmission electron microscope images of the silicates. Left: The originally condensed silicate grains with amorphous structure. Middle: Silicate grains annealed to 500°C still characterized by an amorphous internal structure. Right: The crystallized silicate sample produced at 900°C showing the typical lattice fringes caused by crystalline structures.

At lower temperatures, thermal annealing causes the coalescence of small grains without changing the amorphous structure. The annealing of the originally condensed silicate grains leads to an increase of the grain sizes and decrease of the porosity. The average grain size increases from about 6 nm in the originally condensed silicates to about 14 nm in the 500°C and up to 30 nm in the 900°C annealed sample, whereas the corresponding porosities decrease from 90 to about 65 and 40%, respectively. The silicate sample annealed at 900°C is characterized by randomly oriented enstatite crystallites.

In Figure 6, we present temperature dependences between 200 and 300 K of the amount of remaining water (in percent of the deposited amount) in different samples. Important results related to Figure 6 are the following:

(i) Crystalline silicates (silicates annealed at 900°C) do not trap water molecules. Trapped water has not been detected at 200, 250 and 300 K.

(ii) Silicates annealed at 500°C trap water molecules less efficiently compared to amorphous silicates. Trapped water has not been detected at 250 and 300 K.

(iii) Amorphous silicates mixed with the kerogen material trap water molecules less efficiently compared to amorphous silicates. More kerogen material is produced on silicates; less $H_2O$ molecules are trapped. Trapped water has not been detected at 300 K in both samples and at 250 K in the sample containing more kerogen material.

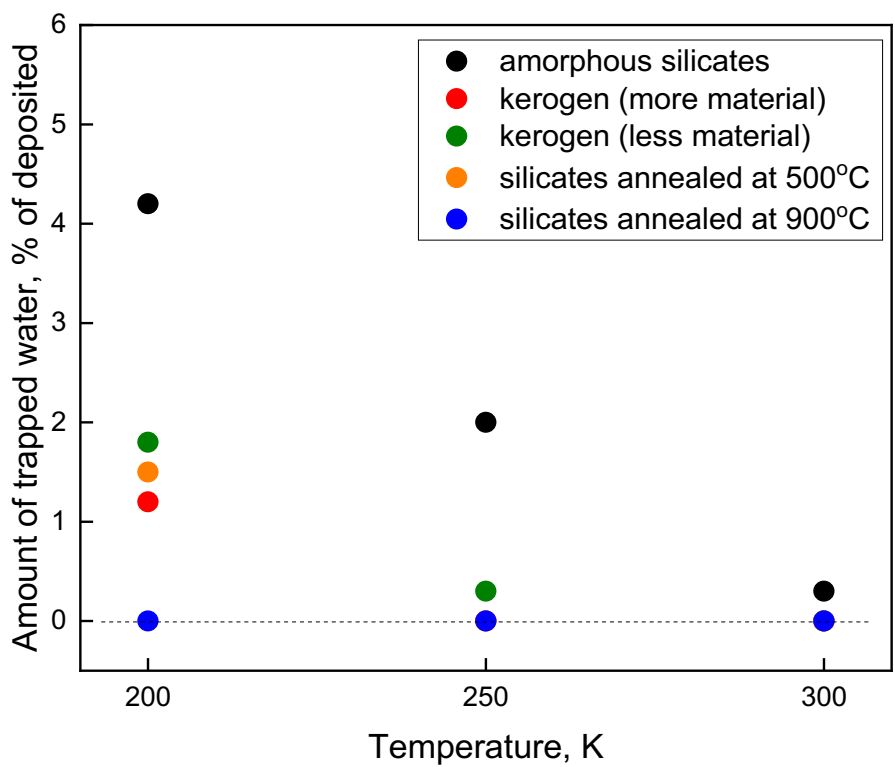

Figure 6. Dependences of the amount of remaining water on various surfaces as a percentage to the deposited amount. Black – amorphous silicates, red and green – amorphous silicates covered by kerogen (red – more material), orange – amorphous silicates annealed at 500°C, blue – crystalline silicates (amorphous silicates annealed at 900°C). Note that the kerogen amounts are not known and only relative amounts were obtained comparing the intensities of CH bands. Note that the 300 K value for kerogen (less material) and the 250 and 300 K values for kerogen (more material) and silicates annealed at 500°C is 0.

## 4. Discussion

**The nature and efficiency of trapping**

Complete desorption of trapped water at about 470 K (~200°C) speaks for the physical nature of its trapping (physisorption rather than chemisorption). Chemisorbed water on amorphous silicates starts to desorb at temperatures of about 700 K and the process of dehydroxylation is completed only at temperatures above the glass transition temperature (Iler 1979; Zhuravlev 2000; Thi et al. 2020). Silica as well as silicates are known to have hydrophilic and hydrophobic surface groups. Physisorption of water on multicomponent silicates is more complex compared to silica because of structural differences. For silica, the available physorption sites are silanol groups of different types. For complex silicates, non-brigding oxygen (NBO) or metal ions on the surface can increase the adsorption efficiency of molecular water. Magnesium silicates containing Mg as a network modifier that can form non-bridging oxygen and Mg ions are efficient adsorbers of water molecules. In addition, OH can also act as a modifier and produce NBOs leading to a change of the structure (Stevenson & Novak 2011).

Thus, the hydrophilic properties of the silicate surface provide the first explanation for the phenomenon of $H_2O$ trapping. This conclusion is in agreement with the results of the calculations on the interaction of water with silicates showing the presence of stronger adsorption sites on the silicate surface (Stimpfl et al. 2006; Muralidharan et al. 2008; King et al. 2010). It is also known that silica becomes less hydrophilic under thermal annealing (Bergna & Roberts 2006). This leads to a gradual decrease of the amount of physisorbed water in silica colloidal crystals with increase of the annealing temperature (Gallego-Gomez et al. 2011; Gallego-Gomez et al. 2012). This explanation is applicable to our results presented in Figure 5, where less efficient trapping is observed for silicates annealed at 500°C and no trapping is observed for silicates annealed at 900°C.

However, trapping efficiency is defined not only by the hydrophilic properties of the silicate surface but also by its porosity and surface area. It is known that for materials with pore size distributions in the range between 2 and 60 nm, a synergistic effect of surface hydroxylation and pore size leads to a strong increase in the water adsorption (Iler 1979; Saliba et al. 2016). In our case, the increase of grain sizes and decrease of porosity with temperature (Figure 6 and the discussion around) clearly play a role as leading to a decrease of the surface area of silicate grains and corresponding decrease of the amount of binding sites.

In the case of amorphous silicates covered by the kerogen-like material, we assume that kerogen partly covers the silicate surface and simply closes binding sites in which water molecules can be trapped (strongly bound). In addition, the kerogen-like material also closes pores. More kerogen material on the silicates leads to higher surface coverage, lower amounts of binding sites, decreased porosity, and, consequently, lower numbers of trapped water molecules.

Thus, the number of water molecules trapped on silicates should have a saturation level as being limited by the number of binding (trapping) sites and presents a tool for estimation of the number of trapping sites. This is an interesting question for further investigations.

**Oxygen depletion problem**

The depletion of elemental oxygen in the ISM is a long standing problem (Jenkins 2009; Whittet 2010; Poteet et al. 2015). A detailed interpretation of observations clearly shows that oxygen disappears from the gas phase faster than it can be explained by its incorporation into refractory silicates and oxides or into ices on the surface of dust grains. An amount of interstellar O missed along the lines of sight increases with the increasing number density of the ISM corresponding to the transition from diffuse to dense phases. As much as a third of the total elemental O budget is unaccounted for in any observed form at the transition between diffuse and dense phases of the ISM and as much as a half is missed in dense phases (Whittet 2010). These value can be corrected using new solar oxygen abundance measurements (see, e.g., (Pietrow et al. 2023) and references therein), however, on the level of a few percent.

One explanation could be that oxygen is trapped with another equally abundant and reactive species, such as carbon and hydrogen, in some solid phase (Jenkins 2009; Whittet 2010). However, to date, there are little if any observational evidence to support this hypothesis (Poteet et al. 2015; Jones 2016). The possible existence of large, greater than 1 μm in diameter interstellar grains (opaque for IR light) containing $H_2O$ ice and, thus, serving as O-reservoirs has been proposed as another explanation (Jenkins 2009). This idea was supported by a number of theoretical models showing grain size distributions in the ISM up to the micrometer range (Clayton et al. 2003; Wang et al. 2014; Mattsson 2016), by investigation of the composition of interstellar grains along the line of sight toward the bright star Ophiuchi (Poteet et al. 2015) giving a tentative evidence for large grains with the radius of about 2.8 μm, and by the results on the rapid formation of dust in the dense circumstellar medium of the supernova 2010jl (Gall et al. 2014), where the extinction curve was fitted by a power-law grain size distribution with minimum and maximum grain radiuses in the interval 0.001 - 5.0 μm. However, the amounts of such large grains and "hidden" $H_2O$ ice are under question.

Water trapped on silicates presents an alternative (or additional) explanation for the oxygen depletion problem. A first attempt to detect trapped water in the diffuse ISM, with its OH stretching band blue-shifted with respect to pure $H_2O$ ice, has been done only recently using ISO and Spitzer data (Potapov et al. 2021). Missed elemental oxygen in the ISM can be explained by much higher abundance of solid-state water that presents a pure trapped water on silicates in the diffuse ISM and a mixture of water ice and trapped water in the dense ISM. The estimates made in (Potapov et al. 2021) for one diffuse region showed that the amount of trapped water detected corresponds to ~2% of the cosmic oxygen budget, ~7% of the oxygen in dust and ~5% of the missing oxygen in the dense ISM. However, statistics is required. To detect trapped water in various diffuse interstellar clouds is one of the goals of the JWST Cycle 1 project "Illuminating the dust properties in the diffuse ISM with JWST". The data were recently obtained and are presently analyzed.

**Origin of water on Earth**

One of the key aspects to reveal the possibility of life on a rocky planet is to understand the delivery of water to its surface. Several scenarios are discussed. The reader can find details in review papers (van Dishoeck et al. 2014; Meech & Raymond 2019; Öberg & Bergin 2021). Assuming that solid-state water exists only beyond the snowline, in the so-called dry scenario, the terrestrial planets are initially built up from planetesimals/pebbles inside the snowline with low (or no) water mass fractions and water is delivered to their surfaces by water-rich asteroids or comets formed outside the snowline. Alternatively, in the so-called wet scenario, the planets either accreted a water-rich atmosphere or formed beyond the snowline. Different variations are being discussed to reconcile the latter idea with the typical inner location of rocky Earth-like planets, either by a subsequent migration of the planet to the terrestrial planet zone or via a time-variable location of the snowline. Here, the transport of pebbles and smaller dust grains should be mentioned, as an importance mechanism for enriching the water content of the inner disk region (Banzatti et al. 2023; Gasman et al. 2023; Grant et al. 2023; Kamp et al. 2023; Perotti et al. 2023). However, one more possibility of the wet scenario is that local planetesimals/pebbles in the time of the Earth formation retained some water at high temperatures through its physisorption or chemisorption on silicate grains. Thus, to understand whether water can be present in the solid state inside the snowline in planet-forming disks is the key question for choosing between the different scenarios.

Silicate grains containing trapped water that forms in cold environments through silicate-ice mixing and survived up to the temperatures of about 470 K present great candidates for wet building blocks of Earth and terrestrial planets. To detect trapped water inside the snowline in planet-forming disks and to quantify its amount is the primary goal of the JWST Cycle 1 project "Solid-state water inside the snowline in planet-forming disks: exploring the origin of water on terrestrial planets". The data were recently obtained and are presently analyzed.

**Formation of planets**

The efficiency of the formation of rocky and giant planets depends on the amount of solid-state material present at the formation location (Thommes & Duncan 2006; Min et al. 2011). Enhanced dust densities due to the presence of ice outside of the snowline are assumed to play an important role in the formation of giant planets. In addition, according to coagulation models (Wada et al. 2009; Wettlaufer 2010) and collisional experiments (Gundlach & Blum 2015; Schräpler et al. 2022), ice-coated grains may stick together much more efficiently, which may lead to more efficient planetesimals/pebbles formation. However, the efficiency of dust coagulation in protoplanetary disks strongly depends on the amount of solid-state water in dust grains (Kimura et al. 2020). Rocky planets accrete their mass mainly through solid-state material from dust grains, pebbles or planetesimals. Thus, evidence for the presence of solid-state water on grains inside the snowline and an estimation of its abundance would lead to a more complete understanding of the formation process of rocky planets. Our laboratory results suggest that trapped water may exist on silicate grains up to the temperatures of about 470 K well covering the rocky planet formation zone.

5. **Conclusions**

The results of a laboratory study of the phenomenon of trapping (strong bonding) of water molecules by silicates in the temperature range of 200 – 600 K are presented. Water molecules can be trapped on silicates up to the temperatures of about 470 K. The efficiency of trapping is strongly dependent on the properties and composition of the surface. The nature of trapping is defined to be physical rather than chemical. The results of the study have important implications to various astrophysically relevant cases, such as the oxygen depletion problem, the origin of water on Earth, and the formation of rocky planets, and motivated new observational studies with JWST, the results of which are expected in the nearest future.

**Acknowledgments**

This study was supported by the Federal Ministry for Economic Affairs and Climate Action on the basis of a decision by the German Bundestag (the German Aerospace Center project 50OR2215). AP acknowledges support from the Deutsche Forschungsgemeinschaft (Heisenberg grant PO 1542/7-1). CJ and HM acknowledge support by the Research Unit FOR 2285 "Debris Disks in Planetary Systems" of the Deutsche Forschungsgemeinschaft (grants JA 2107/3-2 and MU 1164/9-2). TH acknowledges support from the European Research Council under the Horizon 2020 Framework Program via the ERC Advanced Grant Origins 83 24 28.